# Using Polyvinyl Alcohol as Polymeric Adhesive to Enhance the Water Stability of Soil and its Performance

Chunyan Cao,[1] Lingyu Zhao,[2,*] and Gang Li [2,*]

1 School of Electrics and Computer Engineering, Nanfang College, Guangzhou, Guangzhou 510970, China

2 Department of Materials Science and Engineering, Southern University of Science and Technology, Shenzhen 518055, China

*Corresponding e-mail: zhaoly@sustech.edu.cn and 11930690@mail.sustech.edu.cn

**Abstract**

Soil degradation threatens agricultural productivity and food supply, leading to hunger issues in some developing regions. To address this challenge, we developed a low-cost, highly efficient, and long-term stable soil improvement method. We chose polyvinyl alcohol (PVA), a commercially available polymer that is safe and non-degradable, to serve as a soil adhesive. We mixed PVA solution into the soil and applied a drying treatment to enhance the bonding between PVA and the soil, achieving highly water-stable soil. This PVA-stabilized soil exhibits low bulk density, high porosity, and high permeability, making it an ideal substrate for planting. In a germination test, the PVA-stabilized soil revealed a higher germination rate and growth rate compared to those of the non-treated soil. We believe this simple and efficient soil improvement method can restore degraded soil and contribute to sustainable agriculture.

**Key words**: Polymer adhesive, Soil degradation, water-stable aggregate, Sustainable agriculture.

**Introduction**

Agriculture is one of the cornerstones of human society, feeding eight billion people on Earth. In agriculture, soil plays a crucial role.[1] It serves as the substrate for crop growth, providing the necessary nutrients and moisture for plants to thrive.[2] However, improper agricultural practices such as over-tillage, over-fertilization, and excessive use of chemicals can lead to soil degradation. Soil degradation results in decreased land productivity and serious issues like soil erosion, posing significant threats to the sustainability of human society.[3]

Soil aggregates are the basic units of soil structure.[1] Enhancing the stability of soil aggregates is pivotal for improving the performance of degraded soils and restoring their functionality. The formation of soil aggregates in natural soils typically involves interactions among biological, physical, and chemical processes. Soil microorganisms decompose organic matter, yielding colloids and gelatinous substances that act as binding agents, facilitating the cohesion of soil particles into aggregates. However, in soils subjected to improper agriculture practices, these organic binding agents may not be effectively replenished, leading to a decline in aggregate stability and soil performance. Strategies for enhancing soil aggregate stability include the addition of organic matter and polymer amendments. Incorporating organic matter such as compost, manure, or crop residues can increase soil organic matter content, enhance soil microbial activity, and promote the formation and stabilization of aggregates.[4] Nevertheless, the effectiveness of organic matter addition may require extended periods to manifest visibly, and there is a risk of introducing external pathogens or weed seeds. Another approach involves the application of polymer amendments, such as polyacrylamide (PAM),[5, 6] however, its efficacy is limited and it may suffer from a short duration of action.[7] Consequently, there remains a lack of efficient, rapidly effective, and long-term effective soil stabilization methods.

In this study, we propose an efficient, rapid, and long-term stable method for enhancing soil aggregate stability and subsequently improving soil performance. We introduced a polyvinyl alcohol (PVA) into the soil (at a PVA concentration of 0.1% of the soil) through mixing aqueous PVA solution with soil, followed by complete drying.

The large surface area of clay particles in the soil can irreversibly adsorb PVA onto the soil,[8, 9] and the drying treatments greatly enhance this interaction. Through this treatment, PVA acts as a soil binder, significantly enhancing the water stability of soil aggregates. Soil aggregates bonded with PVA do not disintegrate in wet sieving tests and remain stable even after 30 wet-dry cycles and a one-year field experiment without significant degradation. Compared to the original soil, the PVA treated soil shows higher germination rates, highlighting its superiority over the original soil. Our method offers promising commercial prospects due to its low material cost, rapid effectiveness, low energy consumption, long duration, and yield-enhancing capabilities.

**Results and discussion**

**Design principle**

Soil aggregate stability is an important indicator of soil health.[10] Unstable aggregates disintegrate under the impact of rainfall, forming smaller particles that can clog existing surface pores, leading to the formation of crusts. This, in turn, reduces soil permeability and aeration (Figure 1A). At this point, operations such as plowing may seem necessary (Figure 1B). However, plowing not only fails to solve the problem of crust formation permanently but also disturbs the soil further and could potentially make it less healthy. Additionally, lower aggregate stability can lead to surface soil erosion under rainfall.[11, 12]

The goal of this study is to create a soil composed of water-stable aggregates, thereby fundamentally preventing the occurrence of crust formation and compaction (Figure 1C).[13] By improving the water stability of the aggregates, they are better able to withstand the impact of rainfall while maintaining good permeability. This stable soil, with its stable porosity, can also decrease the frequency of tillage (Figure 1D).

Based on considerations of safety,[14, 15] biodegradability,[16-18] and adhesion,[8, 9] we selected polyvinyl alcohol (PVA) as the soil adhesive (Supplementary Text 1). We enhanced the bonding effect through a process of mixing and drying. In this study, we investigated three types of soil: soil from the author's hometown of Chaozhou, referred to as CZ-soil; yellow-soil from the Loess Plateau (from Shanxi province); and red-soil from South China (from Fujian province). The latter two are widely distributed soils in

China and are generally considered to have poor performance, requiring improvement. The samples treated with PVA are referred to as PVA-CZ-soil, PVA-yellow-soil, and PVA-red-soil, respectively.

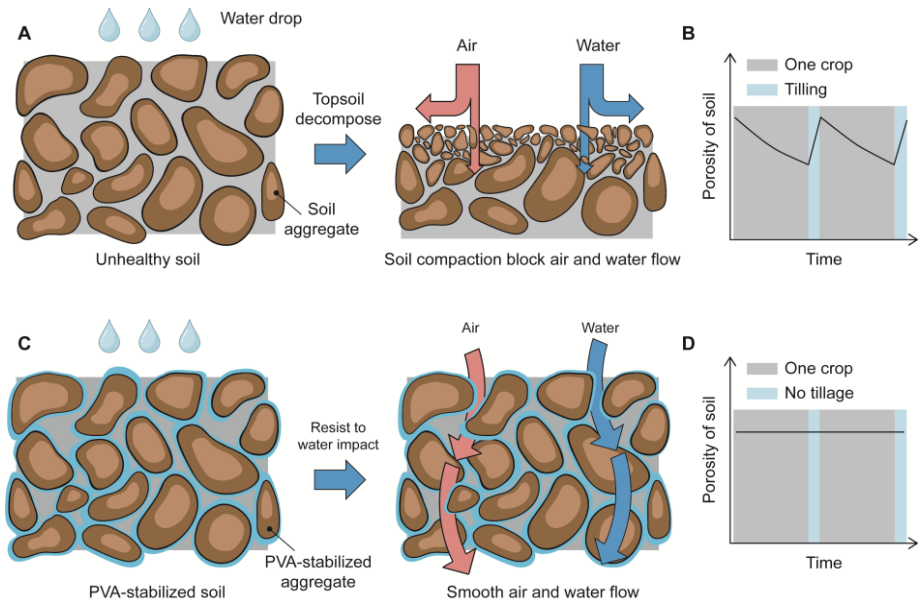

**Figure 1. Schematic illustration of the application of PVA to stabilize soil and endow it with high performance.** A) Unhealthy soil with poor stability disintegrates under water impact, forming small aggregates that can clog surface pores and hinder the flow of air and water. B) Over time, unhealthy soil experiences a decrease in porosity, necessitating periodic tillage to restore pore spaces. C) This study uses PVA as a soil binder to improve the stability of soil, making it stable under water impact and maintaining excellent air and water permeability in the wet state. D) PVA-stabilized soil maintains a stable porosity over time, reducing the need for tillage and labor requirements.

**Mechanical performance**

The mechanical properties of soil are closely related to the amount of binder it contains. As shown in Figure 2A, when dry soil that lacks an organic binder is immersed in water, the interactions between soil particles are disrupted by the water, leading to soil disintegration. However, when PVA is added, the irreversible adhesion of PVA to soil particles can maintain the shape of the soil in water (Figure 2B). In this regard, a higher content of PVA will impart greater mechanical strength to the soil, especially under wet conditions.

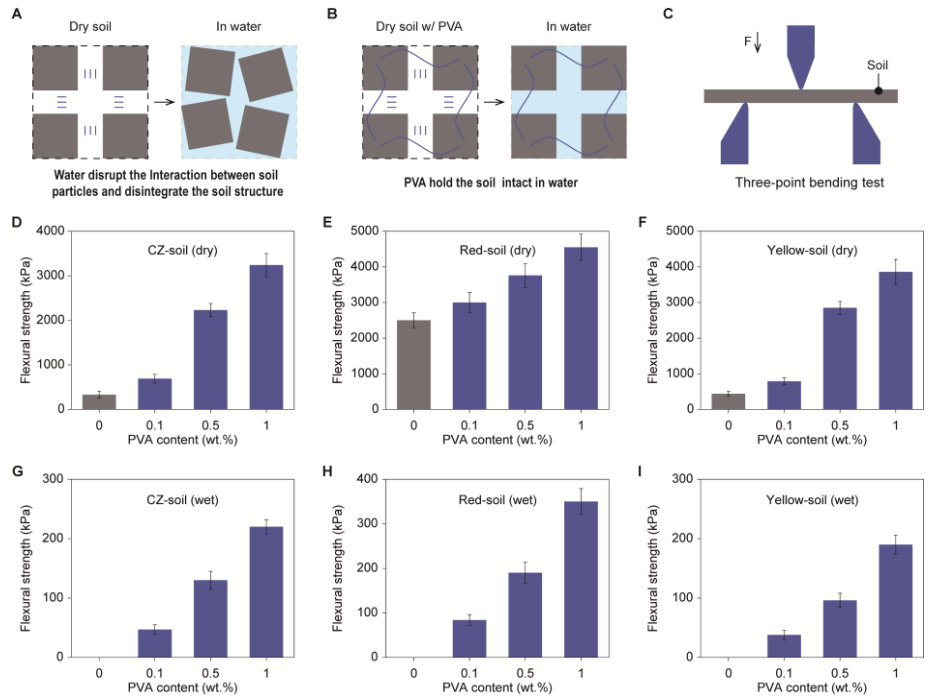

**Figure 2. Mechanical performance of soil with and without PVA.** A) Soil lacking an organic binder tends to disintegrate in water as water disrupts the interaction between adjacent soil particles. B) PVA adheres to the soil, making it intact in water. C) Illustration of the setup for a three-point bending test. D, E, and F) Flexural strength of CZ-soil (D), red-soil (E), and yellow-soil (F) at the dry state with different contents of PVA. G, H, and I) Flexural strength of CZ-soil (G), red-soil (H), and yellow-soil (I) at the wet state with different contents of PVA. Soils without PVA are broken in water, and their flexural strengths are considered as zero.

We conducted three-point bending tests (Figure 2C) to assess the mechanical properties of soils with different PVA concentrations. Taking CZ-soil as an example, the original soil strength was ≈ 330 kPa. After adding 0.1% PVA, the strength increased to ≈ 1000 kPa (Figure 2D). When the PVA content was increased to 1%, the strength further increased to ≈ 3500 kPa (Figure 2D). The same trend was observed for dried red-soil (Figure 2E) and yellow-soil (Figure 2F).

The addition of PVA imparts superior mechanical properties to soils in wet conditions. In the absence of PVA, CZ-soil, yellow-soil, and red-soil all disintegrate in water, as shown in Figure S1. We thus regard their flexural strength as zero. However, with the addition of PVA, these three types of soil can maintain their initial shape in water, as shown in the photos in Figure S2. In three-point bending tests, with a PVA content of 0.1%, the flexural strength of wet PVA-CZ-soil is approximately 50 kPa, and this strength increases to above 200 kPa when the PVA content is further increased to 1% (Figure 2G). Similar improvements in flexural strength can be observed for samples of red-soil and yellow-soil after being moistened (Figure 2H and I). These results reveal that the addition of 0.1% PVA endows the soil with measurable flexural strength in a wet state, indicating a fundamental change in soil stability that further leads to profound changes in other physical properties as well.

**Stability of soil in water**

We investigated the changes in soil due to PVA during wet-dry cycles. We dried samples of CZ-soil containing 0.1% PVA and those without PVA, and sieved them to obtain particles in the size range of 0.5-2 mm for study. During drying, there was no significant difference between the two types of soil (Figure 3A). However, upon adding water, the aggregates of PVA-free CZ-soil disintegrated, whereas the PVA-CZ-soil maintained the same appearance as when it was dry (Figure 3B). This phenomenon was also observed in yellow-soil and red-soil (Figure S3).

We used computed tomography (CT) to analyze the structure of CZ-soil in water. The PVA-free CZ soil had a large amount of aggregates estimated to be less than 0.1 mm in size (Figure 3C), which is much smaller than their initial size range of 0.5-2 mm. Meanwhile, PVA-CZ-soil retained large aggregates in water, with no signs of aggregate breakdown (Figure 3D). The PVA-CZ-soil had more voids, which are beneficial for maintaining soil moisture and high permeability. After wet-dry cycles, the surface of PVA-free soil became denser due to the tendency of small aggregates to compact densely (Figure 3E). On the other hand, the sample with PVA remained rough and porous (Figure 3F).

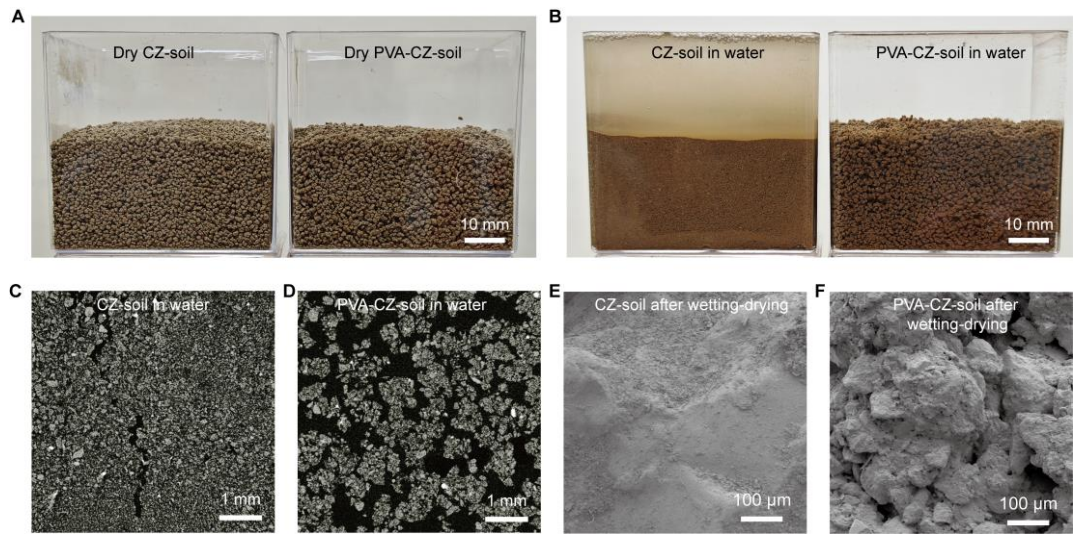

**Figure 3. Comparison of CZ-soil with and without PVA under dry-wet-dry cycles.** A) Photo of CZ-soil and PVA-CZ-soil with sizes ranging from 0.5-2 mm. B) CZ-soil disintegrates in water, while PVA-CZ-soil remains stable. C) A CT slice reveals that CZ-soil has a major amount of aggregates with sizes lower than 0.1 mm and contains a minimal amount of pores. D) PVA-CZ-soil shows aggregates with sizes ranging from 0.5-2 mm in water and provides a higher ratio of pores. E) The surface of CZ-soil is dense after being re-dried from a wet state. F) The surface of PVA-CZ-soil is rough and porous after being re-dried.

We evaluated the water stability of soil improved by PVA using the wet sieving method.[10] Soil samples with and without 0.1% PVA, with particle sizes ranging from 0.5 to 1 mm, were subjected to wet sieving using a 0.25 mm sieve, and the changes in mass after wet sieving were recorded. As shown in Figure 4A, the mass reduction of CZ-soil without PVA after one wet sieving exceeded 70%, while the one with only 0.1% PVA showed a mass change of about 1%. After multiple wet-dry cycles, the stability of soil in wet sieving did not show a significant decrease. For instance, the soil after 30 wet-dry cycles only shows a reduction of 1% in the wet sieving test (Figure 4B). We conducted the same wet sieving on soil after one year of cultivation in the field, and the mass reduction was also below 5% (Figure 4C). The high water stability of PVA-improved soil is mainly attributed to its strong binding with clay particles in the soil, while the low degradation rate of PVA in the soil ensures its long-term stability. Such long-term stability make PVA a much better soil binder than the classical PAM, which efficiency decay in a few months.

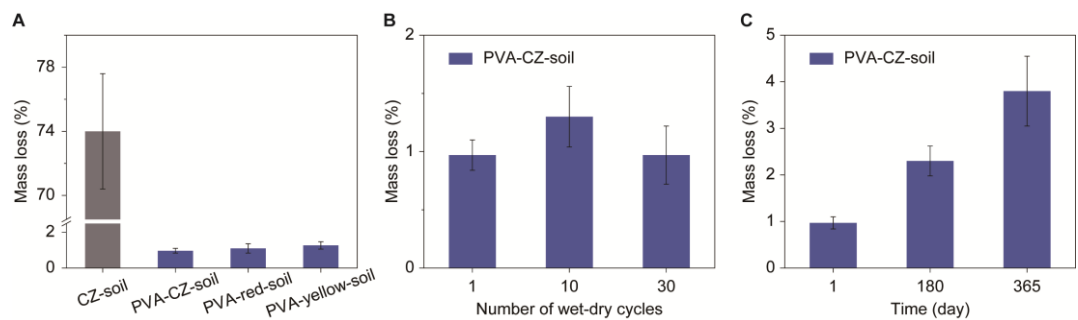

**Figure 4. Water stability of soil with and without PVA in wet sieving tests.** A) The mass loss of CZ-soil and PVA-CZ-soil, PVA-red-soil, and PVA-yellow-soil after one wet sieving test. B) Mass loss of PVA-CZ-soil in wet sieving tests over different numbers of wet-dry cycles. C) Mass loss of PVA-CZ-soil in wet sieving tests over a duration of one year. The soil samples used in the tests have a size ranging from 0.5-2 mm before the test, and all PVA-soils contain 0.1% of PVA.

**Physical properties**

The significant improvement in water stability has greatly enhanced the performance of the soil. We compared the physical properties of the three type of soils with and without 0.1% PVA. The bulk density of the original CZ-soil was 1.2 g cm$^{-3}$ (Figure 5A), with corresponding porosities of 54% (Figure 5B). After PVA amendment, the bulk density of the soil was 0.79 g cm$^{-3}$ (Figure 5A), resulting in a calculated porosity of 70% (Figure 5B), which is 16% higher than that of the original samples. This reduction in bulk density and increase in porosity is also observed in the case of red-soil and yellow soil (Figure S4 A-D).

PVA treatment significantly improves soil permeability. The saturated hydraulic conductivity of soil without PVA was $6 \times 10^{-5}$ cm s$^{-1}$. After PVA amendment, the PVA-CZ-soil exhibits a high hydraulic conductivity of 0.4 cm s$^{-1}$, which is four orders of magnitude higher than that of the original soil (Figure 5C). The high hydraulic conductivity of CZ-soil is due to its small particle size and low porosity, while PVA-CZ-soil maintains larger pores even when submerged in water (Figure 3C), greatly facilitating water permeation. The hydraulic conductivity of both red-soil and yellow-soil can also be enhanced by this PVA treatment (Figure S4E-F). The low hydraulic conductivity of the original soil can lead to surface water accumulation, which not only hinders air diffusion into the soil but also causes runoff and soil erosion. This is one of the reasons for soil erosion in the Loess Plateau. The addition of PVA significantly increases soil permeability, preventing surface water runoff and erosion. Moreover, its high water stability also helps to prevent soil erosion. The method of using PVA-stabilized soil has significant implications for soil and water conservation in regions such as the Loess Plateau.

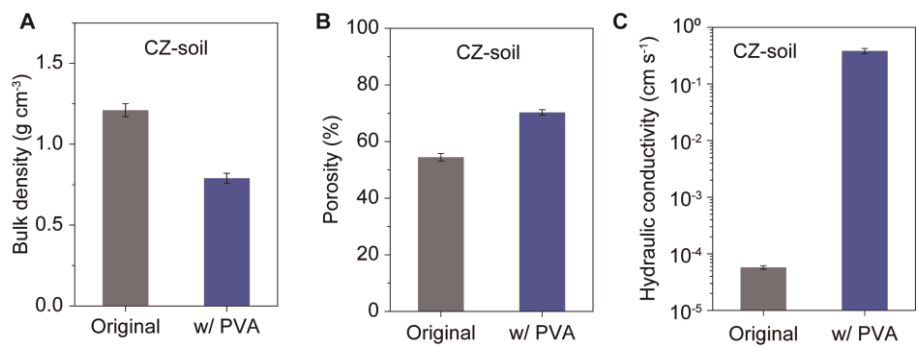

**Figure 5. Changes in physical properties of CZ-soil after PVA treatment.** A) The bulk density of CZ-soil decreases after treatment with PVA. B) The porosity of CZ-soil increases due to the PVA treatment. C) The hydraulic conductivity of PVA-CZ-soil exhibits significant improvement compared to the original sample. The original CZ-soil is the one that passes through a 0.05 mm sieve after being collected from the site. The soil with PVA has a aggregate size ranging from 0.5-2 mm.

**Application in seedling cultivation**

The treatment with PVA significantly enhances the soil's performance in cultivation. We compared the performance of PVA-CZ-soil with CZ-soil in seedling cultivation, with pumpkin being the study object. As shown in Figure 6A, pumpkin seeds were planted in a 12-cell seedling tray, with half using PVA-CZ-soil and the other half using CZ-soil. In a typical scenario, pumpkin seeds emerged from PVA-CZ-soil after two days (Figure 6A). Pumpkin seedlings grew well in PVA-CZ-soil after seven days, whereas in this typical test, no pumpkin seed succeeded in germinating from the CZ-soil (Figure 6B). We counted the germination rate over six experiments and the results showed that 60% more pumpkins were germinated from the PVA-CZ-soil compared to CZ-soil (Figure 6C). We removed the soil from the roots of the pumpkin seeds on day 2 and photographed them as shown in Figure 6D. Pumpkin roots in CZ-soil had only one main root and the root system was short and sparse, while those grown in PVA-CZ-soil had five long roots with many secondary roots. The mass of the root system in the ones cultivated in PVA-CZ-soil was about twice that in CZ-soil (Figure 6E), indicating a preferred condition was provided by the PVA-CZ-soil. The strong root system supported strong growth. On day seven, the mass of the shoot of the pumpkins in PVA-CZ-soil was higher than in CZ-soil (Figure 6E), further demonstrating the beneficial effects of PVA-CZ-soil on plant growth.

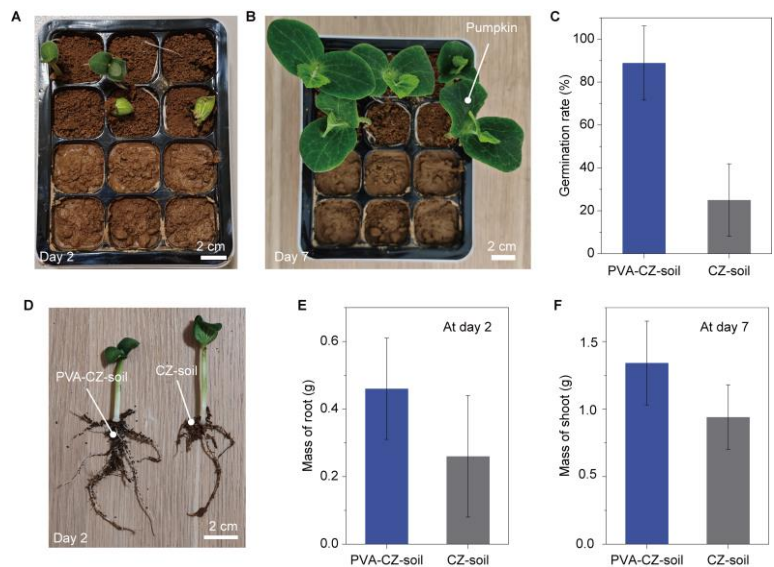

**Figure 6. Application of PVA-CZ-soil in seedling cultivation.** A) Pumpkin seeds germinated in a 12-cell seedling tray after two days when planted in PVA-CZ-soil and CZ-soil. B) Pumpkin plants grew well in PVA-CZ-soil after 7 days, whereas no germination occurred in CZ-soil. C) Comparison of the germination rate of pumpkins in PVA-CZ-soil and CZ-soil. Results are from six experiments. D) The roots of pumpkin seedlings on day 2. E) The root mass of pumpkin seedlings grown in PVA-CZ-soil and CZ-soil on day 2. F) The shoot mass of pumpkin seedlings grown in PVA-CZ-soil and CZ-soil after 7 days.

In addition to pumpkin seedling cultivation, we have explored the use of PVA-CZ-soil, PVA-red-soil, and PVA-yellow-soil in various qualitative tests. This includes the cultivation of corn (Figure S5) and mini-cyclamen (Figure S6). We have also experimented with growing a multitude of species in a glass tank using PVA-red-soil, an endeavor that would have been unattainable with native soil (Figure S7). We believe that these PVA-soils hold great promise not only in agriculture but also excel as superior substrates for potted plants in horticulture.

**Comparison with existing methods**

We perform a comparative analysis of our PVA-based soil enhancement method against other soil amelioration techniques that utilize compost, biochar, or PAM. These three categories of materials are the popular choices for soil modification. Our evaluation primarily concentrates on the material costs, as well as the time and energy investments associated with each method.

Composting is a process that requires a significant time investment, typically taking several months to mature.[19] Once prepared, it then needs an additional extended period to exert a beneficial effect on the soil, as its potency relies on microbial decomposition.[20] In this regard, composting is a time-consuming process. In contrast, our method begins to exert its effects as soon as the soil-PVA mixture is dried. Under sunny conditions or in dry weather, the time required for our method is minimal, usually amounting to just one week, and often less than that—perhaps two or three days. Furthermore, PVA solutions are commercially available, which further enhances the time-efficiency of our approach.

Biochar is another type of soil amendment, but its production often requires high-temperature conditions, which can be energy-intensive.[21, 22] In contrast, our PVA solution is easily obtainable from commercial sources and requires minimal energy for mixing with soil. Additionally, the drying process is naturally aided by sunlight or wind. Furthermore, biochar shares similar issues with compost, in that its effectiveness is cumulative and it also needs a long time to reach its full potential.

PAM is the most extensively researched polymer soil conditioner, offering a unique advantage in terms of lower material costs. In comparison, compost or biochar typically require application in much larger quantities, often tons per hectare, while PAM is applied in kilogram amounts. However, PAM is suffered by its relatively low efficiency and a rapid decline in effectiveness over time. The material cost of our PVA method may be comparable to PAM's if we opt to focus on modifying only the topsoil, the materials usage is about ten kilogram per hectare.

Our PVA-based soil stabilization method offers significant benefits in terms of time, energy, and material savings. Additionally, it provides the added advantages of simplicity, long-term effectiveness, and tolerance to various additives, thereby expanding its potential applications. The mixing and drying process is straightforward and can be further simplified by adopting a spray-drying method, particularly when focusing on topsoil improvement. With respect to stability over time, our experimental results have shown that the PVA-treated soil remains stable with no observable decline after one year, demonstrating its exceptional long-term stability. This stability is a result of the strong interaction between PVA and the soil, as well as the absence of PVA-degrading microorganisms in soil. This long-term stability reduces the overall input of PVA into the soil over a longer period. Our PVA method hold high tolerance to other additive. PVA functions by binding the clay particles in the soil, and the addition of materials such as compost, biochar, and sawdust does not interfere with PVA's efficacy. Our method is compatible with other soil improvement techniques, opening the door to a wider range of soil modification possibilities.

**Potentials**

To focus on the efficacy of PVA, the prior discussion has concentrated solely on its interaction with soil, excluding other substances or additives to prevent overcomplication. Thus, the discussion serves as a proof of concept, leaving significant room for future enhancements and optimizations for actual application.

The PVA-soil mixture can be further refined by integrating organic materials such as straw or sawdust, as well as inorganic materials like vermiculite and perlite. The inclusion of light organic matter can reduce bulk density (Figure S8A), while the porous inorganic material can enhance porosity (Figure S8B), both of which are instrumental in improving soil properties.

The mixing and drying process can be optimized by a spraying-drying process. Field experiments have demonstrated that spraying the plowed soil with a PVA solution, at a rate of 10 grams of PVA per square meter, is adequate to stabilize the topsoil.

This PVA-based soil improvement method holds promise for horticultural applications, where natural peat is commonly used, often leading to wetland destruction. Replacing peat with PVA-soil in potting has shown to be effective (Figure S6 and S7).

Our method can be employed to mitigate soil erosion caused by water. By enhancing soil stability in water and increasing hydraulic conductivity, it minimizes soil detachment and runoff, protecting the soil from erosion by rainfall.

PVA-stabilized soil may be more suitable for sponge city initiatives, signified by the tunable and high hydraulic conductivity. The large the size of the aggregate, the greater the hydraulic conductivity (Figure S9). The highly permeable PVA-soils is thus the preferable sponge, having the potential to alleviate the negative impacts of heavy rain in urban areas.

**Conclusion**

The irreversible absorption of PVA on clay particles of soil renders it an excellent soil binder, capable of creating water-stable soil. Soils treated with 0.1% PVA are able to withstand sustained flexural stress in water and exhibit minimal mass loss during wet sieving tests, properties that are not present in the original samples. The soil formed by water-stable aggregates bonded by PVA demonstrates lower bulk density, higher porosity, and higher hydraulic conductivity compared to the original soils, indicating a significant enhancement in physical soil properties. In a small-scale planting experiment, PVA-treated soils exhibited a higher germination rate than untreated soils. In summary, using PVA as a soil binder is a low-technology-requiring and highly efficient method. Its advantages include rapid action, low material cost, minimal energy consumption, and long-term reliability, making it an ideal choice for improving soil performance.

**Materials and methods**

**Chemicals and soils:** Polyvinyl alcohol (KURARAY POVAL™ 60-98, 98-99% mol% hydrolysis) was purchased from a commercial source. The first soil were collected from

Chaozhou city, Guangdong province. The second soil is yellow soil from Shanxi province, and the third type is red soil from Fujian province, which were both purchased from Taobao.com. All soils were dried and passed through a 0.05 mm sieve to remove large particles of sand. Tap water was used throughout the entire experiment.

**Fabrication of PVA-soil:** Ten grams of PVA powder is added to 90 grams of water and stirred in a 95 °C water bath until completely dissolved to obtain the PVA solution. Then, the PVA solution, dried soil, and water are mixed together in a composition such that the mass ratio of PVA to soil is 0.1% and the mass ratio of water to soil is 25%. The mixture is thoroughly mixed and then drying in an atmosphere.

**Mechanical test of soil:** Dry soil was mixed with water and different amounts of PVA solution, with the water content in the mixture set at 25%. The mixture was then transferred to a mold (20 cm in length, 3 cm in both width and thickness) and dried in the atmosphere. The flexural strength of the dried specimen was tested using a universal testing machine (XLD-20E, Jingkong Mechanical Testing Co., Ltd) with a three-point bending setup. For the wet specimen, the soil sample was soaked in water for 1 hour before the test.

**Wet sieving test:** Dried soil and PVA-soil with particle sizes ranging from 0.5-2 mm are collected. These soils are placed above the sieve with a mesh size of 0.25 mm. Then, the sieve is immersed in a water bath and moved up and down with a displacement of 3 cm and a frequency of 30 times per minute for a duration of 10 minutes. After this process, the sample remaining on the sieve is collected and dried again. The change in mass before and after the sieving process is used to indicate the water stability of the soils.

**Measurement of particle density:** Ten gram of dried soil is placed into a 50-mL pycnometer, then 20 g water is added into the bottle. The bottle is kept in vacuum for 1 hour to remove bubbles. Finally the bottle is transfer to atmosphere and filled with water. The mass of the bottle, soil and water is weighed as $m_{bws}$. The same pycnometer only filled with water is weighed as $m_{bw}$. The particle density of soil, $\rho_P$ is determined as:

$$\rho_P = \frac{m_s}{m_s + m_{bw} - m_{bws}} \times \rho_w$$

Where $m_s$ is the mass of PVA-soil and $\rho_w$ is the density of water at 25 °C (since the experiment is conducted at room temperature).

**Measurement of bulk density and porosity:** The sieved PVA-soil are filled into a plastic collum with volume of 1 L. Oscillation and percussion is applied to enhance package. Then the PVA-soil is weighed and the bulk density $\rho_b$ is calculated by dividing mass to volume. The porosity (*f*) of PVA-soil is determined as:

$$f = 1 - \frac{\rho_b}{\rho_p}$$

**Measurement of hydraulic conductivity:** The PVA-treated soil was placed in a cylindrical apparatus with a cross-sectional area of 30 cm². Under constant head conditions and a steady vertical water flow, the volume of water passing through the specimen was measured over a specific time interval. Subsequently, the hydraulic conductivity of the specimen was calculated using Darcy's law.

For untreated soil, hydraulic conductivity is measured by a falling head condition.

**Pumpkin seedling cultivation:** The pumpkin seeds are from Daziran Agricultural Technology Co., Ltd. in Shouguang City. Before sowing, the seeds are soaked in water at 40 degrees Celsius for 1 hour, then covered with a moist paper for 24 hours. After that, the seeds are placed in wet soil and covered with 0.5-1 cm of soil. The seedlings are then placed in a bright environment, and the lid of the seedling tray is covered for the first four days. The temperature during the experiment ranges from 20 to 30 °C. Each of the 12-cell seedling trays has six cells filled with PVA-CZ-soil and six cells with CZ-soil. Each seedling tray counts as one trial. The number of germinated seeds on the seventh day is used to calculate the germination rate. On the second day, the germinated pumpkins are carefully removed from the soil, and the soil is washed off the roots with water and then the moisture is removed with paper towels. The roots are then trimmed and weighed. Six random pumpkin seedlings are selected for root mass measurement. On the seventh day, the pumpkins are separated from the soil, the roots are carefully trimmed off, and the weight of six pumpkin shoots is measured to assess the growth rate of pumpkins in different soils.

**Acknowledgments:** This work was funded by the Young Innovative Talents Project of Universities in Guangdong Province, China (No. 2022KQNCX137), National Natural Science Foundation of China (No. 52103301), Natural Science Foundation of Guangdong Province (No. 2023A1515012835), Basic Research Program of Shenzhen (No. JCYJ20230807093559046).

**Competing interests:** The authors declare that they have no conflicts of interest.

**Data and materials availability**: All data are available in the main text or the supporting information.

**Author contribution:** Conceptualization, G.L.; Methodology, G.L.; Investigation, C.C., Y,Z, and G.L.; Writing – Original Draft Preparation, G.L.; Writing – Review & Editing, C.C. and L.Z.; Supervision, L.Z.; Project Administration, G.L.; Funding Acquisition, C.C. and L.Z.

**Supporting information:** Figure S1 to S9.

Supporting Information

# Using Polyvinyl Alcohol as Polymeric Adhesive to Enhance the Water Stability of Soil and its Performance

Chunyan Cao,[1] Lingyu Zhao,[2,*] and Gang Li [2,*]

1 School of Electrics and Computer Engineering, Nanfang College, Guangzhou, Guangzhou 510970, China

2 Department of Materials Science and Engineering, Southern University of Science and Technology, Shenzhen 518055, China

*Corresponding e-mail: zhaoly@sustech.edu.cn and 11930690@mail.sustech.edu.cn

Supplemental text 1, Properties of PVA

1.1. Safety of PVA. PVA is non-toxic and oral edible.[1]

1.2. Biodegradability of PVA in soil. Previous studies indicate that PVA is poorly degradable in soil,[2-6] primarily due to the scarcity of PVA-degrading microorganisms. In one study, PVA sheets were buried in 18 natural soil sites with varying compositions and climates for two years.[5] Their results revealed that the degraded PVA was less than 10%. This poor degradability ensures the long-term usage of PVA in soil. The second reason for poor biodegradability is attributed to the irreversible absorption of PVA on clay.[3] Another research shows that the interaction between PVA and clay can hinder the attack of organisms on PVA.[7] Only 4% of clay-absorbed PVA degraded in a solution containing PVA-degrading organisms after one month of incubation.[7] In the contrast, 34% of the free PVA was degraded.[7] This result implies that our PVA-soil can sustain for a long period even with PVA-degrading microorganisms present.

1.3. Adhesion of PVA on soil. The PVA can be irreversibly absorbed onto the clay particles in soil.[7, 8] This irreversible absorption between PVA and clay can be regarded as PVA strongly adhere on clay particles. Afterall, the introduction of PVA enhances the interaction among the clay particles and thus the whole soil, since the slit and sand will be surrounded by those clay and PVA.

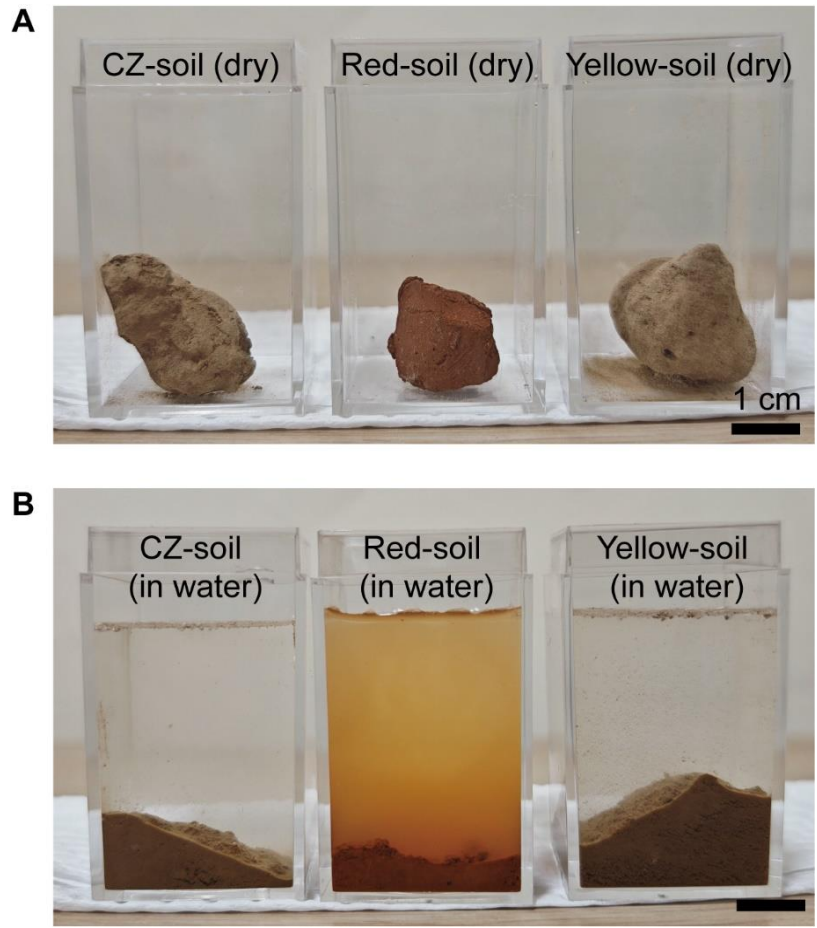

**Figure S1. Soil disintegration in water.** The large grain of soil (A) breaks down into water and loses its initial shape (B).

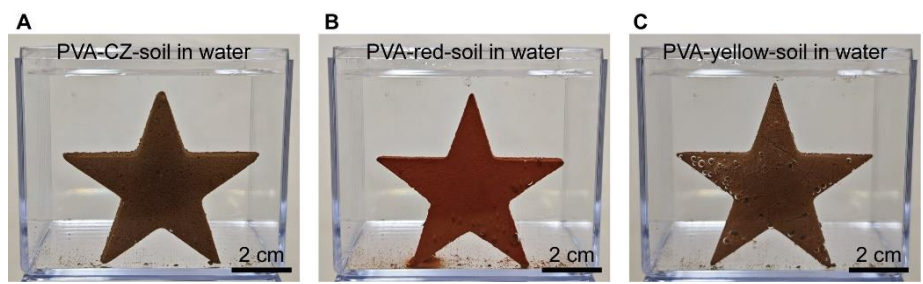

**Figure S2. PVA-soil in water.** The PVA-CZ soil (A), PVA-red-soil (B), and PVA-yellow-soil, with a star shape, show integrity in water, demonstrating their remarkable stability in water.

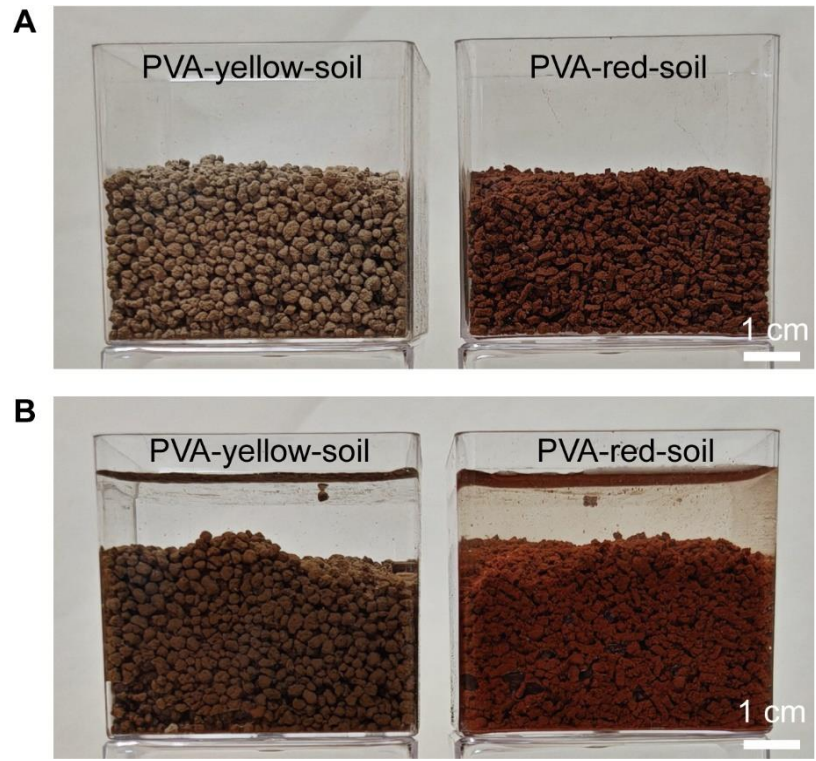

**Figure S3. PVA-soil in dry state and wet state.** Transferring from a dry environment (A) to an underwater condition does not break down the aggregate of PVA-soil (B).

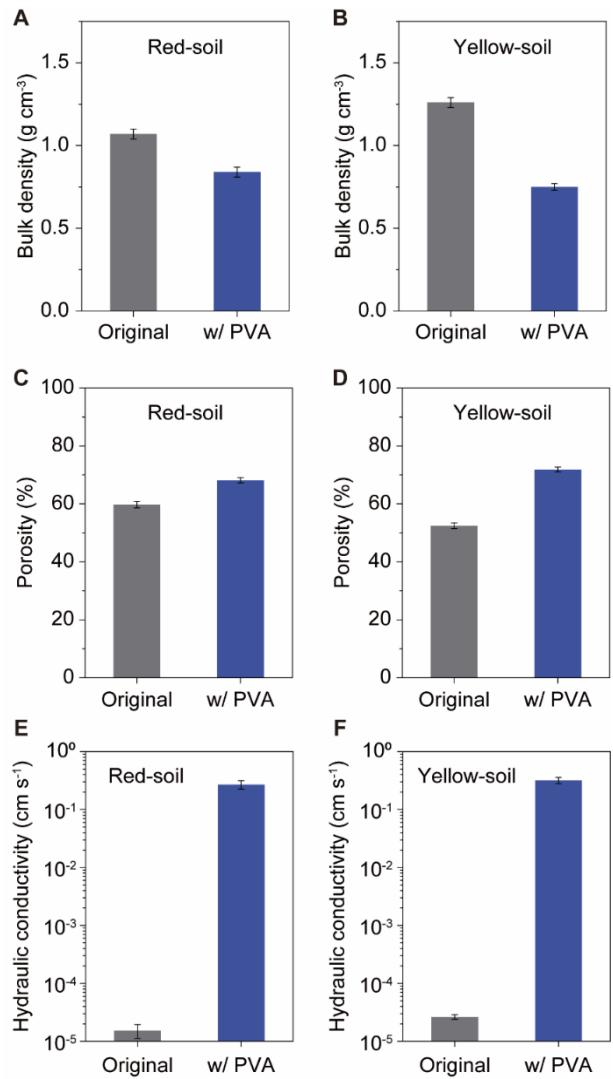

**Figure S4. Comparison of physical properties of soil with and without the addition of PVA**. A) Bulk density of red-soil. B) Bulk density of yellow-soil. C) Porosity of red-soil. D) Porosity of yellow-soil. E) Hydraulic conductivity of red-soil. F) Hydraulic conductivity of yellow-soil.

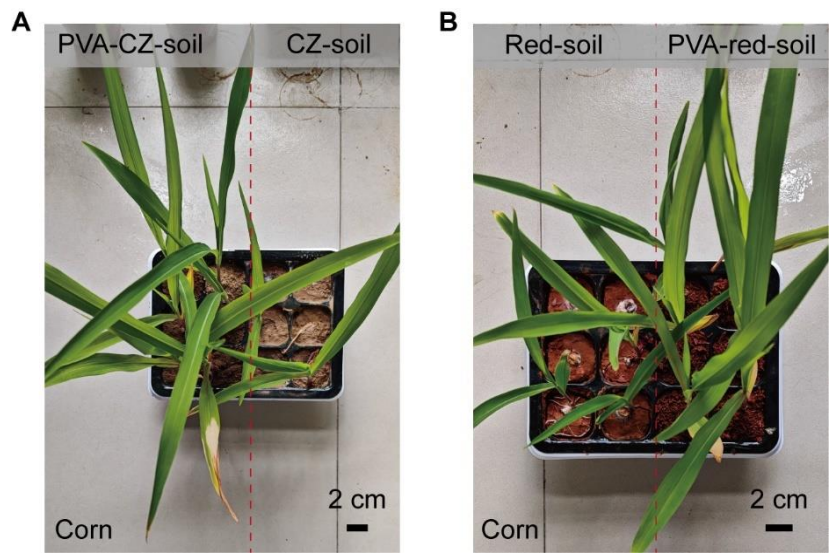

**Figure S5. Using soil containing PVA and soil without PVA for corn cultivation.** a) Corn seedlings grow well and all germinate in PVA-CZ-soil, but only 1/3 of the seedlings germinate and have weaker growth in CZ-soil. b) Corn seedlings grow well and all germinate in PVA-red-soil, but only half of the seedlings germinate in red-soil. Photos are recorded at day 14 after planting.

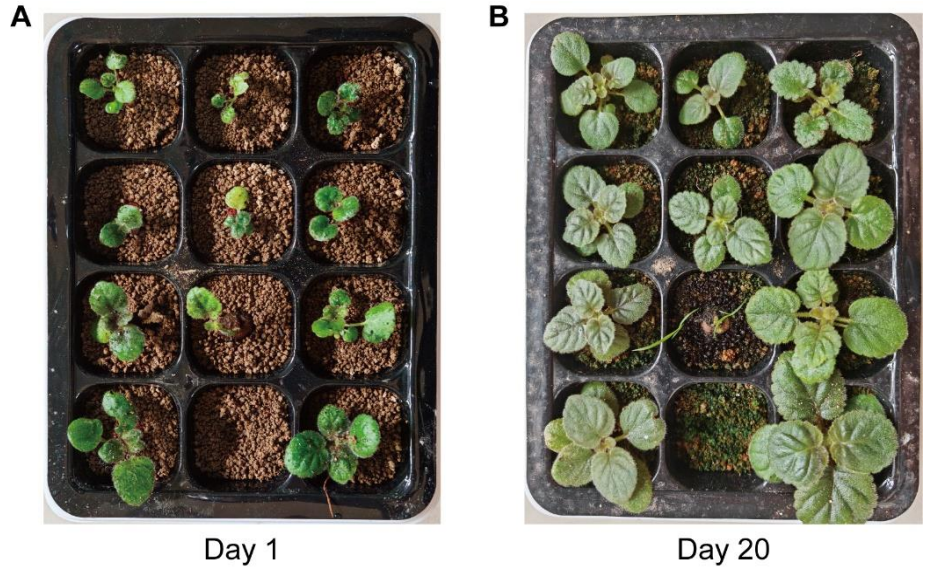

**Figure S6. Growth of mini-cyclamen in PVA-yellow-soil.** A) Mini-Cyclamen just transplanted into PVA-yellow-soil. B) Growth of Mini-Cyclamen after 20 days, with 10 out of 11 plants growing well.

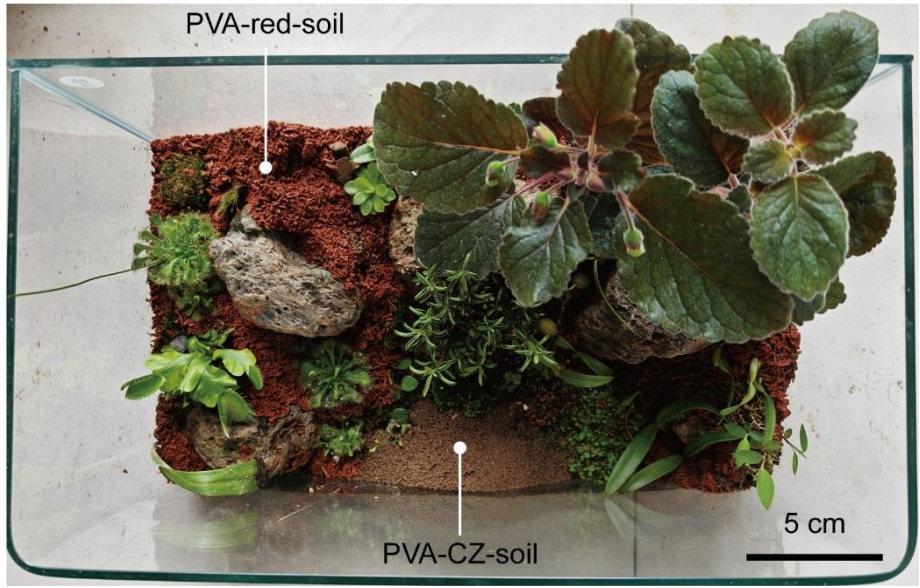

**Figure S7. Plants grow in the PVA-red-soil and PVA-CZ-soil.**

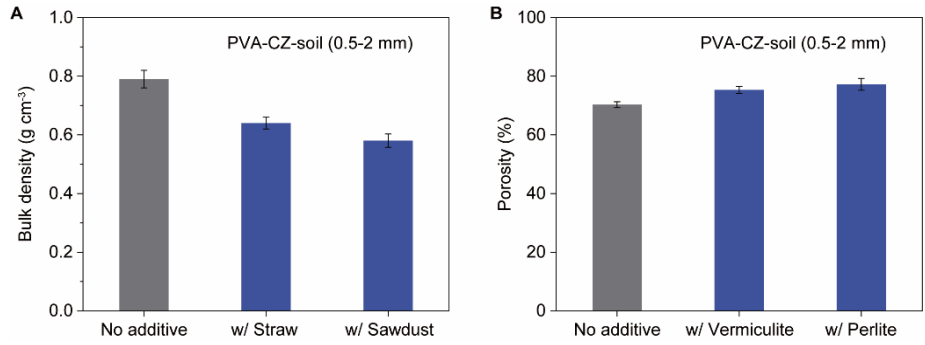

**Figure S8. Comparison of physical properties of PVA-CZ-soil with and without the additive.** A) The addition of 20 wt.% rice straw or sawdust reduces the bulk density of PVA-CZ-soil because rice straw and sawdust are much lighter than soil. In this experiment, straw and sawdust are approximately 2 mm in length. B) The inclusion of 10 vol% vermiculite and perlite increases the porosity of PVA-CZ-soil because these inorganic materials are porous. In this experiment, vermiculite and perlite are roughly 1 mm in size.

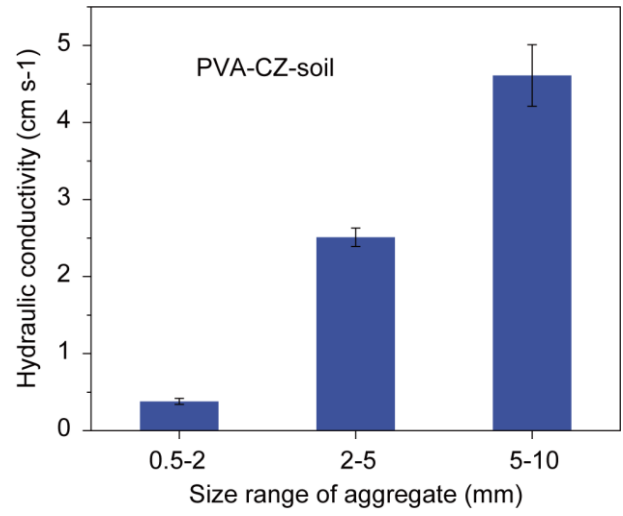

**Figure S9. Hydraulic conductivity of PVA-CZ-soil with different aggregate sizes.**